\documentclass[journal]{IEEEtran}
\usepackage[utf8x]{inputenc}
\usepackage{pdfpages}
\usepackage{cite}
\usepackage{subfigure}
\usepackage{amsmath}
\usepackage{braket}
\usepackage{listings} 
\usepackage{hyperref}
\usepackage{morefloats}
\usepackage{lipsum}
\usepackage{dcolumn}
\usepackage{array}
\usepackage{eqnarray}
\usepackage{fixltx2e}
\usepackage[normalem]{ulem}
\usepackage{cuted}
\usepackage{balance}

\begin{document}               

%\markboth{Kerem Y. Camsari, Samiran Ganguly, Deepanjan Datta, Supriyo Datta}
%{Combining Coherent Spin Currents and Nano-Magnets}
%
%\makeatletter
%\def\bbordermatrix#1{\begingroup \m@th
  %\@tempdima 4.75\p@
  %\setbox\z@\vbox{%
    %\def\cr{\crcr\noalign{\kern2\p@\global\let\cr\endline}}%
    %\ialign{$##$\hfil\kern2\p@\kern\@tempdima&\thinspace\hfil$##$\hfil
      %&&\quad\hfil$##$\hfil\crcr
      %\omit\strut\hfil\crcr\noalign{\kern-\baselineskip}%
      %#1\crcr\omit\strut\cr}}%
  %\setbox\tw@\vbox{\unvcopy\z@\global\setbox\@ne\lastbox}%
  %\setbox\tw@\hbox{\unhbox\@ne\unskip\global\setbox\@ne\lastbox}%
  %\setbox\tw@\hbox{$\kern\wd\@ne\kern-\@tempdima\left[\kern-\wd\@ne
    %\global\setbox\@ne\vbox{\box\@ne\kern2\p@}%
    %\vcenter{\kern-\ht\@ne\unvbox\z@\kern-\baselineskip}\,\right]$}%
  %\null\;\vbox{\kern\ht\@ne\box\tw@}\endgroup}
%\makeatother

\title{Non-Equilibrium Green's Function based Circuit Models for Coherent Spin Devices}
%\title{Modular Approach to Spintronics}
\author{\IEEEauthorblockN{Kerem~Y.~Camsari\IEEEauthorrefmark{1},
        Samiran~Ganguly\IEEEauthorrefmark{2},
        Deepanjan~Datta\IEEEauthorrefmark{3},
				and~Supriyo~Datta\IEEEauthorrefmark{1}~\IEEEmembership{Fellow, IEEE}}\\
\IEEEauthorblockA{\IEEEauthorrefmark{1}School of Electrical and Computer Engineering, Purdue 
University, West Lafayette, IN 47907\\ \{kcamsari,datta\}@purdue.edu\\
				\IEEEauthorrefmark{2} Dept. of Electrical and Computer Engineering, University of Virginia, Charlottesville, VA 22904\\ sganguly@virginia.edu\\
				\IEEEauthorrefmark{3}GLOBALFOUNDRIES, Bangalore, INDIA }
}
\maketitle

\begin{abstract}
With recent developments in spintronics, it is now possible to envision ``spin-driven''  devices with magnets and interconnects that require a new class of transport models using generalized Fermi functions and currents, each with four components: one for charge and three for spin. The corresponding impedance elements are not pure numbers but $4\times4$ matrices. Starting from the Non-Equilibrium Green's Function (NEGF) formalism in the elastic, phase-coherent transport regime, we develop spin generalized Landauer-B\"uttiker formulas involving such $4\times 4$ conductances, for multi-terminal devices in the presence of Normal-Metal (NM) leads. In addition to usual ``terminal'' conductances describing currents at the contacts, we provide ``spin-transfer torque'' conductances describing the spin currents absorbed by ferromagnetic (FM) regions inside the conductor, specifying both of these currents in terms of Fermi functions at the terminals. We derive universal sum rules and reciprocity relations that would be obeyed by such matrix conductances. Finally, we apply our formulation to two example Hamiltonians describing the Rashba and the Hanle effect in 2D. Our results allows the use of pure quantum transport models as building blocks in constructing  circuit models for complex spintronic and nano-magnetic structures and devices for simulation in SPICE-like simulators.  
\end{abstract}
%\pacs{}
\maketitle

\section{Introduction}
\label{sec:intro}

The emergent field of spintronics has grown at a rapid rate over past three decades, starting from low temperature experiments in metallic magnetic structures to commercialized  memory chips that are considered as the future of embedded and consumer markets \cite{bhatti2017spintronics}. The field keeps forging ahead with developments of novel materials and phenomena driven primarily by advances in abilities to manipulate materials at the nanoscale. 

To enable analysis and modeling of such diverse materials and phenomena, modular, circuit techniques generalized to account for spin-currents have been proposed   \cite{brataas_non-collinear_2006,rychkov2009spin,behin-aein_switching_2011,manipatruni_modeling_2012, srinivasan_modeling_2013, camsari2015modular}. Such ``spin-circuits'' explicitly account for the transport of  spin-currents through channels and interfaces allowing the combination of a diverse range of materials to analyze new functional devices and experimental structures. 

In this paper we show how phase-coherent materials and devices can be systematically expressed to be included within the existing spin-circuit framework starting from the Non-Equilibrium Green's Function (NEGF) formalism \cite{datta_electronic_1997},  a widely used method for quantum transport. The objective of this work is to demonstrate how  a fully phase-coherent channel can be recast as an electrical circuit and combined with spin-circuits that are obtained from different theoretical methods, such as spin-diffusion equations. 

\subsection{Development of spin-circuits: A brief overview}
\label{sec:mod_hist}

Development of  multi-component spin-circuits  can be traced back to the successful application of the 
2-Current Model \cite{PhysRevLett.21.1190} for the analysis of the Current-Perpendicular-to-Plane Giant Magneto-Resistance (CPP GMR) devices in the collinear configuration in which the two contact magnets are either parallel or anti-parallel to each other. This approach was later extended into a 4-Current theory to treat non-collinear spin-currents derived from the quantum mechanical density matrices to relate  charge and spin currents to their respective electrochemical potentials \cite{brataas_non-collinear_2006,rychkov2009spin}. Subsequently, the 4-current theory was expressed in terms of a general class of 4-component circuits that can be seamlessly integrated to SPICE-like circuit analysis tools  \cite{behin-aein_switching_2011,srinivasan_modeling_2013}. Additionally,  spin-transport equations were integrated with magnetization dynamics using the stochastic Landau-Lifshitz-Gilbert Equation to incorporate time dependent effects to be solved self-consistently with underlying transport equations. This approach was shown to be amenable for SPICE-like circuit analysis \cite{manipatruni_modeling_2012} and further extended and compounded in  Ref. \cite{camsari2015modular} along with an accompanying open source library and example circuit models built using the library which are available from the project's portal\cite{_group:_????}.  With recent developments such as the discovery of the Giant Spin Hall Effect (GSHE) \cite{miron2010current,liu_spin-torque_2012} and with the advent of Topological Insulators\cite{li2014electrical}, the list of such SPICE-compatible spin-circuit modules were extended to include these phenomena \cite{hong2016spin,hong2016spin2} and shown to successfully model complex spintronic devices involving many different modules \cite{camsari2015modular,_group:_????}. The circuit nature of these modules allowed their seamless integration even though different modules were derived from different theoretical methods, such as spin-diffusion equations for GSHE, quantum transport methods for interfaces between ferromagnets and normal metals. 

 This modular, circuit-based approach has  been extended to include new physics such as different types of voltage control of magnetic anisotropy \cite{camsari2018equivalent,torunbalci_modular_2018} and has been used for analyzing emerging spintronic devices \cite{ganguly2016evaluating} or the evaluation of novel computation schemes based on the stochastic behavior of low-barrier nanomagnets \cite{camsari2017stochastic}. 

\subsection{Main motivation}

In this paper we present a systematic formalism to incorporate materials that require a quantum mechanical treatment into the framework of what we call the ``modular approach''\cite{camsari2015modular}, that allows device models built using these materials to be analyzed using standard  circuit simulators like SPICE. In addition to the general results, we  provide two illustrative examples of spin-circuits that are obtained  for two well-known Hamiltonians, one describing a Rashba channel and another describing a Hanle channel both treated in 2D.  The main motivation of this paper is to separate the quantum mechanically coherent regions of heterogeneous devices from regions where a classical Boltzmann or diffusion equation approach is sufficient, but at the same time expressing the coherent regions in terms of generalized conductances that can be combined with the rest of these classical 4-component circuits. This approach could simplify the analysis of complex structures where a full quantum treatment might be intractable and  unnecessary.

\subsection{Coherent Currents: Formal Definition}
\label{sec:coh-cur}

\begin{figure}
\includegraphics[width=\linewidth]{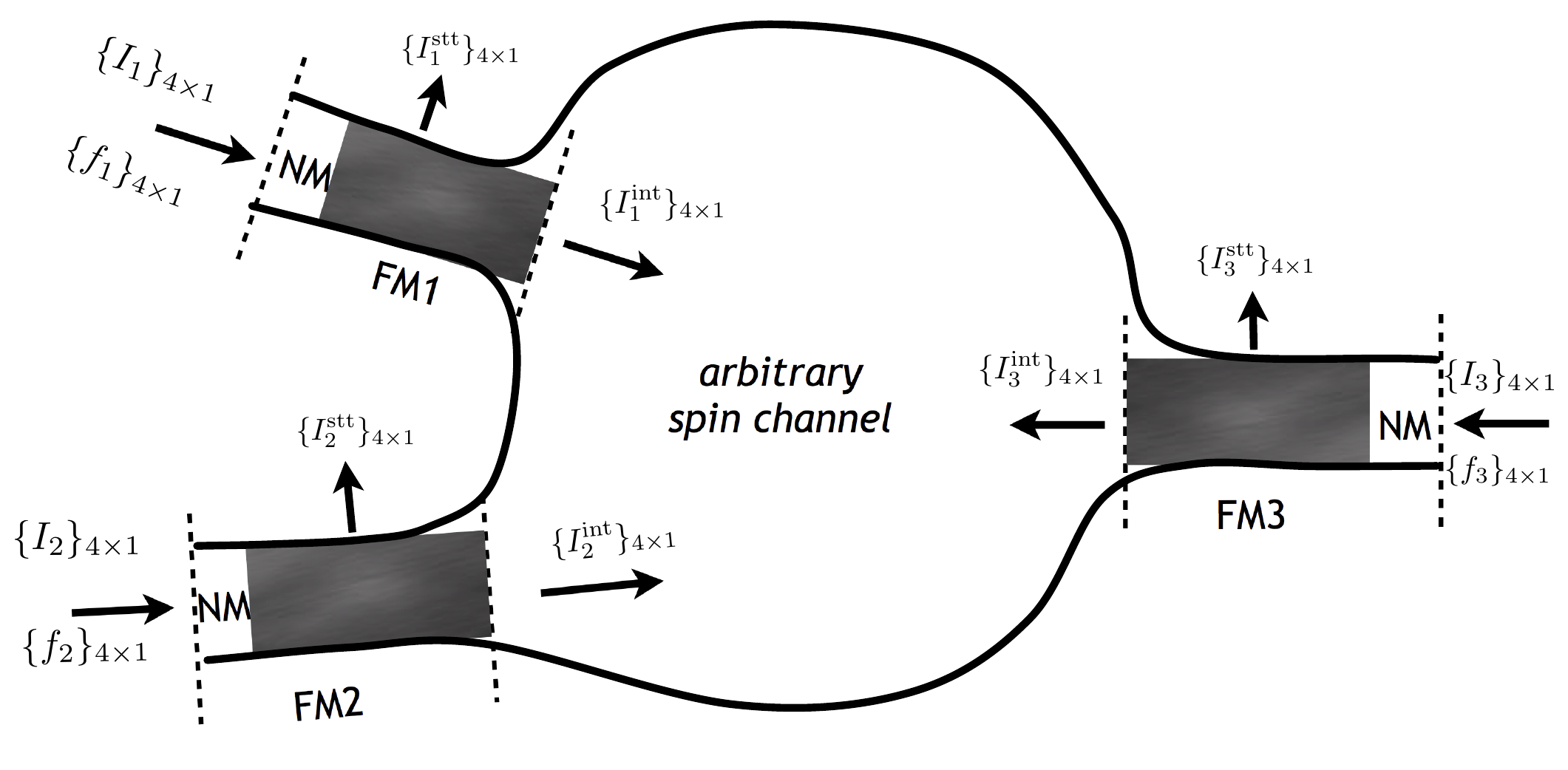}
\caption{Schematic of a multi-terminal conductor described by an arbitrary Hamiltonian that could include spin-orbit coupling and/or time-reversal asymmetry. Each terminal is a Normal Metal (NM)  channel that allows proper definitions of  both charge and spin potentials and currents, described by a four-component quantity. The $(4\times1)$ terminal current is related to the $(4\times1)$ occupation function by a $(4\times4)$ terminal conductance matrix, Eq.~(\ref{eq:1}), and to the $(4\times1)$ spin-transfer-torque current by the spin-transfer-torque conductance matrix, as shown in Eq.~(\ref{eq:2}).}
\label{fi:fig1}
\end{figure}

The Landauer-B\"uttiker equation\cite{buttiker_four-terminal_1986}  relates the terminal currents $I_m$ to the terminal Fermi functions $f_n$ (fig.~(\ref{fi:fig1}))
\begin{subequations}
\label{eq:0}
\begin{equation}
\widetilde{I}_m (E) = \frac{1}{q}\sum_n \widetilde{G}_{mn}(E) f_n(E)
\end{equation}
or to the electrochemical potentials $\mu_n$ for linear response
\begin{equation}
 I_m  = \frac{1}{q}\sum_n G_{mn} \mu_n 
\end{equation}
\end{subequations}
These equations have been widely used to describe phase-coherent and elastic transport in conductors and in view of recent developments in the field of spintronics\cite{liu_spin-torque_2012,liu_current-induced_2012,demidov_magnetic_2012}, it is natural to ask whether Eq.~(\ref{eq:0}) can be extended to describe spin currents and spin potentials in phase-coherent conductors. 

This can be done by defining $(4\times 1)$ currents $\widetilde{I}_m$, Fermi functions $f_m(E)$ and potentials $\mu_m$, each having one charge component and three spin components which are related by $4\times4$ conductance matrices $\widetilde{G}_{mn}$ leading to Landauer-B\"uttiker style expressions of the form:

\begin{equation}
\label{eq:1}
\left\{\widetilde{I}_m (E)\right\}= \frac{1}{q}\sum_n \left[\widetilde{G}_{mn}(E)\right] \left\{f_n(E)\right\}
\end{equation}

In addition to the ``terminal'' conductance matrix defined by Eq.~(\ref{eq:1}), we also need ``spin-transfer-torque'' conductances that relate the terminal potentials to the internal currents absorbed within specified surfaces inside the conductor:
\begin{equation}
\label{eq:2}
\{\widetilde{I}_m (E)\}^{\rm stt}= \frac{1}{q}\sum_n \left[\widetilde{G}_{mn}(E)\right]^{\rm stt} \{f_n(E)\}
\end{equation}
Typically these could be the difference between interface and terminal currents (fig.~(\ref{fi:fig1}),  $I_m^{\rm int}$ and $I_m$ respectively) representing the spin current absorbed by the Ferromagnet (FM). This spin-torque current is required as the input to a separate Landau-Lifshitz-Gilbert (LLG) equation which we will not be addressing in this paper, and can be found in the ref. \cite{camsari2015modular}. 

The main objective of this paper is  to provide Non-Equilibrium Green's Function (NEGF)-based expressions for both the terminal and the spin-transfer-torque conductances of Eq.~(\ref{eq:1}) and Eq.~(\ref{eq:2}). In Section(\ref{sec:negf}), we summarize the state-of-the-art standard NEGF formulation\cite{theodonis_2006_anomalous,PhysRevB.79.214432,kalitsov2009spin,kalitsov2013spin,oh2009bias,tang2010influence,datta_voltage_2012,waldron2006nonlinear,ke2008disorder,chen_spin_2009,mahfouzi_spin-orbit_2012,mahfouzi_charge_2012} which provides a benchmark for all our results. Next, we obtain our central results (Section(\ref{sec:derive})) namely,  Eq.~(\ref{eq:negf}) for terminal conductances and  Eq.~(\ref{eq:int}) for spin-transfer-torque conductances. 

In Section \ref{sec:sum} and \ref{sec:rec} , we show that Eq.~(\ref{eq:negf}) and Eq.~(\ref{eq:int}) automatically satisfy various sum rules and reciprocity relations ensuring charge current conservation, absence of terminal spin currents in equilibrium, and the spin-generalization of Onsager's reciprocity relations, which are all fundamental checks  for a theoretically sound transport formalism. 

In Section \ref{sec:circuit}  we provide a generic 4-component circuit representation that can be used to implement terminal and spin-transfer-torque conductances in SPICE-like simulators. Although the circuit components we show are based on our NEGF-based expressions, the 4-component circuit we provide can be used with different $4\times 4$ conductances derived from other microscopic theories, such as Scattering Theory. 

Finally in Section {\ref{sec:hamtocirc}} we show how 4-component spin-circuits can be obtained directly from starting model Hamiltonians, choosing two  examples, namely the Rashba Effect and the Hanle Effect that have both been observed in various experiments\cite{koo2009control,choi2015electrical,balakrishnan2013colossal} and utilized in spin-logic proposals  \cite{datta1990electronic}.  

We also note that the expressions we provide for these conductances are model-independent and could be used in conjunction with any microscopic Hamiltonian, first principles, tight-binding or otherwise, that we may choose to use to describe the conductor. 

The results presented here can be used in conjunction with existing building blocks in spintronics\cite{camsari2015modular}, bridging models from quantum transport to semi-classical models permitting the analysis of devices that would be too large for a direct quantum transport modeling without losing essential spin physics. 

\section{NEGF Formalism}
\label{sec:negf}

Our starting point for the conductances shown in Eq.~(\ref{eq:1}-\ref{eq:2}) is based on the NEGF(Chapter 8 in  \cite{datta_electronic_1997} and Chapter 19 in  \cite{datta_lessons_2012})  formalism. The main inputs to NEGF are the self-energy functions ($\Sigma$) that  describe the coupling of the channel to the external contacts, and the Hamiltonian describing the channel itself $(H)$. The two central quantities of interest in NEGF, the retarded Green's function ($G^R$) and the electron correlation matrix ($G^n$) are given in terms of these inputs:
\begin{subequations}
\label{eq:NEGF}
\begin{equation}
  {G^R} = {\left[ {EI - H - \Sigma } \right]^{ - 1}} \hfill \\
\label{eq:gr}
\end{equation}
\begin{equation}
G^n = G^R \Sigma^{in} G^A
\label{eq:kel}
\end{equation}
\end{subequations}
%\begin{figure}
%\includegraphics[scale=0.30]{Figure2.eps}
%\caption{(Color online) Schematic of the multi-terminal conductor modeled by the  NEGF method. Hamiltonian matrix, H, models the shaded region while the self-energy matrices, $\Sigma$ model the contacts together with $2\times 2$ Fermi functions serving as boundary conditions. The Hamiltonians of the magnetic regions  of size $2n\times 2n$ embedded in  a zero matrix $2N\times 2N$ are labeled $H^m_{FM}$ are used in the calculation of spin-transfer-torque conductances defined by Eq.~(\ref{eq:2}). Once $G^n$ and $G^R$ are known from Eq.~(\ref{eq:gr}) and Eq.~(\ref{eq:kel}), Eq.~(\ref{eq:op}) can be used to calculate terminal or spin-transfer-torque currents inside the device in the standard NEGF formalism.}
%\label{fi:fig2}
%\end{figure} 
where $\Sigma$ is the sum of all self-energy matrices, $H$ is the device Hamiltonian, $I$ is the $2N \times 2N$ identity matrix, $N$ being the lattice points of the conductor, $E$ is energy and $G^A$ is the advanced Green's function, the Hermitian conjugate of  the retarded Green's function: $G^A= (G^R)^\dagger$.  $\Sigma^{in}$ appearing in Eq.~(\ref{eq:kel})  is the total `inscattering' of electrons that includes the electron injection from the contacts. We are following the notation used in \cite{datta_electronic_1997,datta_lessons_2012} with  $G^n \equiv -i G^<$ and $ \Sigma^{in} = -i \Sigma^<$.

Once the Green's function (Eq.~(\ref{eq:gr})) and the electron correlation   matrix (Eq.~(\ref{eq:kel})) are known, the net flux of spins entering into the conductor  volume ($\Omega$) can be expressed by tracing the NEGF current operator with Pauli spin matrices \cite{datta_electronic_1997,datta_lessons_2012}:
\begin{equation}  \mathrm{tr.}\left[ {{S_\alpha }{I_{op}}} \right] \hspace{-2pt} = \frac{q}{h}{\text{tr}}.\hspace{-6pt}\;\left[ {i \hspace{-2pt}\;{S_\alpha }\hspace{-6pt}\;\left( \begin{gathered}
   {{G^n}H - H{G^n}}  +\hfill \\
   {{{\text{G}}^n}{\Sigma ^{^\dag }} - \Sigma \;{G^n}} + \hfill \\
    {{{\text{G}}^R}{\Sigma ^{in}} - {\Sigma ^{in}}{G^A}}  \hfill \\ 
\end{gathered} \hspace{-6pt}\ \right)}\hspace{-6pt}\ \right]
\label{eq:op}
\end{equation}
where $\Sigma$ and $\Sigma^{in}$ represent the total self-energy and total inscattering matrices, summed over all contacts and $S_\alpha$, an ``expanded" Pauli spin matrix  such that ${S_\alpha } = I \otimes {\sigma _\alpha }$, $I$ being the $N\times N$ identity matrix ($N$: number of lattice points) and $\sigma_\alpha$ is the $2\times2$ Pauli spin matrix for a spin direction $\alpha$, and $I_{2\times 2}$ for charge.

Eq.~(\ref{eq:op}) is widely used in the literature to calculate currents through conventional spin-torque devices, such as MTJs  \cite{theodonis_2006_anomalous,PhysRevB.79.214432,kalitsov2009spin,kalitsov2013spin,oh2009bias,tang2010influence,datta_voltage_2012,waldron2006nonlinear,ke2008disorder,chen_spin_2009}, and through  sophisticated spin-devices {\cite{mahfouzi_spin-orbit_2012,mahfouzi_charge_2012}  for both terminal spin currents, and for internal spin currents involving spin-transfer-torque calculations. Therefore, the main benchmark for our results (Eq.~(\ref{eq:negf})-Eq.~(\ref{eq:int})) has been to make extensive numerical comparisons with Eq.~(\ref{eq:op}), using random Hamiltonians representing arbitrary spin channels to ensure the validity of our expressions presented in the next section.

\section{Derivation of Central Results}
\label{sec:derive}
NEGF matrices defined in Eq.~(\ref{eq:NEGF}) can be specified as a Kronecker product of a real space matrix (of  size $N\times N$) and a spin space component (of size $2\times2$). For example, the ``broadening matrix'' due to non-magnetic contact $m$ can be written as: 
\begin{subequations}
\begin{equation}
\label{eq:gam}
\Gamma_m = i(\Sigma_m - \Sigma^\dagger_m)= \gamma_m \otimes I_{2\times 2}
\end{equation}
where $\gamma_m$ is the $N\times N$ real space component of the full broadening matrix, and it is the anti-Hermitian part of the self-energy matrix $\Sigma_m$. 

When the contacts are driven out of equilibrium by external sources, the inscattering function has to be modified   through its spin component:
\begin{equation}\Sigma _m^{in} =   {\gamma _m} \otimes \left( f^c_m I + \vec{f}_m^s \cdot \vec\sigma\right)
\label{eq:siginf}
\end{equation}
\end{subequations}
Where $(f^c I + \vec{f}^s \cdot \vec \sigma)$ is the $2\times2$ matrix specifying the occupation probabilities of spin and charge components at a given contact which reduces to $f^c I$ for ordinary charge-driven transport.

Next, we observe that the current operator of Eq.~(\ref{eq:op})  is zero at steady state since it represents a) the  sum of all the inflow through the contact boundaries  and b) the ``recombination/generation  (R/G)'' currents  within the conductor volume. The R/G  currents are ordinarily zero for charge currents, however, may exist for spin-currents due to magnetic fields or spin-orbit coupling within the conductor.

In Supplementary Information, we show that the total influx for a given spin direction  $\alpha$ (Eq.~(\ref{eq:op})) can be mathematically decomposed into these two distinct components, the first being the spin-current injected by contact $m$:  \begin{equation}
I_m^\alpha  = \frac{q}{h}{\text{ tr}}.\left(   \begin{gathered}
  {i \ S_\alpha }\left[ {{G^n}\Sigma _m^{^\dag } - {\Sigma _m}{G^n}} \right] \hfill \\
   + i \ S_\alpha \left[  {{G^R}\Sigma _m^{in} - \Sigma _m^{in}{G^A}} \right] \hfill \\ 
\end{gathered}  \right)
\label{eq:inj}
\end{equation}
And  the second component being the generation of spin currents within the conductor volume:
\begin{equation}
S^\alpha_G=\displaystyle \frac{q}{h} \  \mathrm{tr.} \left[ i \  S_\alpha \left( H {G^n} - {G^n} H\right) \right]
\label{eq:gen}
\end{equation}
so that Eq.~(\ref{eq:op}) can be written as: 
\begin{equation}
 \mathrm{tr.}\left[ {{S_\alpha }{I_{op}}} \right] \hspace{-2pt} =\sum_m I_m^\alpha + S^\alpha_G  = 0 
 \end{equation}
\textbf{Terminal Conductances:} Defining the terminal conductances as ${\left[ {{\widetilde{G}_{mn}}} \right]^{\alpha \beta }} = \displaystyle\frac{1}{q}\frac{{\partial {\text{ }}I_m^\alpha }}{{\partial {\text{ }}f_n^\beta }}$ and substituting  Eq.~(\ref{eq:kel}), Eq.~(\ref{eq:siginf}) and Eq.~(\ref{eq:gam}) in Eq.~(\ref{eq:inj}) the conductances can be expressed as:
\begin{equation}
\begin{gathered}
  {\left[ {{\widetilde{G}_{mn}}} \right]^{\alpha \beta }} = \frac{{{q^2}}}{h}{\text{ tr. }}\left[ {i\left( \begin{gathered}
  {S_\beta }{S_\alpha }{G^R}{\Gamma _m} \hfill \\
   - {S_\alpha }{S_\beta }{G^A}{\Gamma _m} \hfill \\ 
\end{gathered}  \right){\delta _{mn}}} \right] \hfill \\  \\
 \quad \quad \quad \quad \ \ \ \  \ \ \ \ \ \ \  - \ {\text{tr. }}\left[ {{\text{ }}{S_\alpha }{\Gamma _m}{G^R}{S_\beta }{\Gamma _n}{G^A}} \right] \hfill \\ 
\end{gathered} 
\label{eq:negf}
\end{equation}
which is the central result for terminal conductances defined in Eq.~(\ref{eq:1}). Eq.~\ref{eq:negf} also seems consistent with a scattering matrix approach outlined in \cite{PhysRevB.84.035412} (See Eq.~(65) in \cite{PhysRevB.84.035412}).  

\textbf{Spin-transfer-torque Conductances:} The spin-transfer-torque absorbed by the FM regions inside the channel are quantified by the negative ``generation'' rate within these volumes: 
\begin{equation}\begin{gathered}
 -S^{FM}_G  = \frac{q}{h}{\text{  tr. }}\left[ i \  {{S_\alpha }\left( {H_{FM}^m{G^n} - {G^n}H_{FM}^m} \right)} \right] \hfill \\ 
\end{gathered}
\label{eq:div}
\end{equation}
where $H_{FM}^m$ is the $2n\times 2n$ Hamiltonian matrix of the ferromagnetic layer (for a magnet with $n$ physical points), embedded in a $2N \times 2N$ zero matrix.  This current can then be used to define spin-transfer-torque conductances that provide the spin-torque absorbed by the FM ${\left[ {\widetilde{G}_{mn}} \right]^{\rm{stt}-\alpha \beta }} = \displaystyle\frac{1}{q}\frac{\partial\left(-S^{FM}_G\right)}{\partial \ f_n^\beta}$ and substituting Eq.~(\ref{eq:kel}) in Eq.~(\ref{eq:div}) we obtain:
\begin{equation}
{\left[ {\widetilde{G}_{mn}} \right]^{\rm{stt}-\alpha \beta }} = \frac{{{q^2}}}{h}{\text{tr.}}\left[ {i\left(\hspace{-3pt}\begin{gathered}
  H_{FM}^m{S_\alpha } \hfill \\
   - {S_\alpha }H_{FM}^m \hfill \\ 
\end{gathered}\hspace{-2pt}\right)\hspace{-2pt}{G^R}{S_\beta }{\Gamma _n}{G^A}} \right]
\label{eq:int}
\end{equation}
which is our central result for $\displaystyle\left[\widetilde{G}_{mn}\displaystyle\right]^{\rm stt}$ defined in Eq.~({\ref{eq:2}). 

We also note that Eq.~(\ref{eq:int}) is general and can be used within any closed surface within the device, such as magnets that are situated in the middle of the device as well as those situated by the contacts. 

\textbf{{Reducing to the Charge Limit:}} The standard result\cite{Meir_Wingreen} for pure charge conductance is a subset of the  terminal conductance matrix shown in Eq.~(\ref{eq:negf}), and can simply be obtained  by using $I$ in place of $S_\alpha$  and $S_\beta$:
\begin{equation}
{\left[ {{\widetilde{G}_{mn}}} \right]^{cc}} = \frac{{{q^2}}}{h}\ {\text{tr. }}\left( {{\Gamma _m} \ A  \ \delta_{mn} - {\Gamma _m}{G^R}{\Gamma _n}{G^A}} \right)
\label{eq:charge}
\end{equation}
where we have made use of the NEGF identity $A = i\left[ {{G^R} - {G^A}} \right]$, A being the `spectral density' matrix per unit energy.
\section{Sum Rules}
\label{sec:sum}
In this section, starting from Eq.~{(\ref{eq:negf} and \ref{eq:int}) we show universal sum rules that the proposed conductance expressions analytically satisfy. We start with the general multi-terminal conductance matrix relating currents and occupation functions at different terminals:
\begin{equation}
\begin{bmatrix} I^c \\ I^z \\ I^x \\ I^y \end{bmatrix}=
\begin{bmatrix}
\widetilde{G}^{cc} & \widetilde G^{cz} & \widetilde G^{cx} & \widetilde G^{cy}  \\               \widetilde G^{zc} & \widetilde G^{zz} & \widetilde G^{zx} & \widetilde G^{zy} \\                \widetilde G^{xc} & \widetilde G^{xz} & \widetilde G^{xx} & \widetilde G^{xy}  \\            \widetilde G^{yc}& \widetilde G^{yz} & \widetilde G^{yx} & \widetilde G^{yy}
\end{bmatrix}
\begin{bmatrix} f^c \\ f^z \\ f^x \\ f^y \end{bmatrix}\nonumber
\end{equation}
where each entry in the conductance matrix is a $P\times P$  matrix while the currents $I^{c,s}$ and occupation functions   $f^{c,s}$  are $P\times 1$ column vectors,  $P$ being the number of terminals in the conductor; so that  the submatrix $\widetilde{G}^{cc}$ can be identified as the conductance matrix describing the coherent charge currents.   \\ \\
\textbf{{Charge Conservation (Terminal):}} Regardless of how charge currents are generated through $\widetilde{G}^{cc}$ or $\widetilde{G}^{cs}$, charge conservation requires them to add up to zero at steady state, in both linear response and high-bias regimes. Linear response is characterized by low input voltages, $(\mu_i - \mu_j)/kT \ll 1$, where $\mu_{i}$ is the electrochemical potential of contact $i$ that determines $f_i$, while high-bias is characterized by high input voltages $(\mu_i - \mu_j)/kT \gg 1$. Charge conservation requires:
\begin{equation}
\sum\limits_{m = 1}^P {\left[\widetilde{G}_{mn}\right]^{c\beta}}  = 0
\label{eq:sum1}
\end{equation}
We show in the appendix that both these equations are analytically satisfied by the conductance matrices  of Eq.~(\ref{eq:negf}).
\\ \\
\textbf{{Equilibrium Currents (Terminal):}} In equilibrium, there are no spin accumulations in the contacts since they are assumed to be non-magnetic, making all occupation functions have the form:  ${f_{eq}} = {\left[ {\begin{array}{*{20}{c}}
  {{f_0}}&0&0&0 \end{array}} \right]^T}$ at a given energy. We then show (Supplementary Information):
 \begin{equation}
\sum\limits_{n = 1}^P {\left[\widetilde{G}_{mn}\right]^{\alpha c}}  = 0
\label{eq:sum3}
\end{equation}
where $\alpha=c$ ensures no net charge current flows through the terminals in equilibrium, whereas $\alpha=s$ ensures the same for spin currents. The latter, once a debated result (See \cite{kiselev_prohibition_2005} and references therein}) was established from Scattering Theory in the context of spin-orbit coupling for devices with non-magnetic leads. In Supplementary Information we prove this result analytically starting from Eq.~(\ref{eq:negf}) and observe that having magnets or magnetic fields (in addition to any spin-orbit interaction) inside the conductor does not change this basic conclusion. We also note that there are no general sum rules for the spin to spin conductances $\left[\widetilde{G}_{mn}\right]^{ss}$.  \\  \\ 
\textbf{{Charge Conservation (Spin-transfer-torque):}} Since no charge currents can be  generated or absorbed within the FM regions,  the generation term $S_G$ becomes identically zero (by choosing $S_\alpha=I$ in Eq.~(\ref{eq:int})) requiring: 
\begin{equation}
\left[ \widetilde{G}_{mn} \right]^{\rm{stt}-c\alpha} = 0
\end{equation}
for all (m,n).

Finally, under equilibrium conditions, the spin-torque applied to the ferromagnetic layer does not necessarily vanish, unlike terminal equilibrium currents, suggesting interesting practical possibilities due to spin currents exerting spin-torque under zero bias conditions.

\section{Reciprocity}
\label{sec:rec}
In this section, starting from time-reversibility conditions for Green's functions we show that the terminal conductances (Eq.~{\ref{eq:negf}}) satisfy the spin-generalization of Onsager's reciprocity in linear response conditions (a result discussed in detail in\cite{jacquod_onsager_2012}): 
\begin{equation}
{\left. {{{\left[ {{\widetilde{G}_{mn}}} \right]}^{\alpha \beta }}} \right|_{ + B}} = {\left. {{{\left[ {{\widetilde{G}_{nm}}} \right]}^{\beta \alpha }}} \right|_{ - B}}{( - 1)^{{n_\alpha } + {n_\beta }}}
\label{eq:ref}
\end{equation}
where the exponents $n_\alpha$ and $n_\beta$ are 0 for charge and 1 for spin indices, respectively. Physically, Eq.~(\ref{eq:ref}) can be justified by noting that reversing time causes spin-currents and spin-voltages to change sign, while charge currents and charge voltages remain invariant \begin{footnote}{The magnetic moment arising due to the spin can be  visualized to arise from a microscopic current loop around the electron, with magnetic moment equal to one Bohr magneton, and reversing time reverses this imaginary current loop.}\end{footnote}, requiring the $n_{\alpha,\beta}$ factors in Eq.~(\ref{eq:ref}).
To show that our conductances satisfy Eq.~(\ref{eq:ref}), we first observe that time-reversibility conditions for Green's functions requires:
\begin{equation}{\left. {{G^R}} \right|_{ + B}} = {\left. {{S_y}{{\left( {{G^R}} \right)}^T}{S_y}} \right|_{ - B}}
\label{eq:time}
\end{equation}
where $G^R$ is the retarded Green's function matrix, $S_y$ is  the expanded Pauli spin matrix in the y-direction while T denotes matrix transpose. In Supplementary Information, we start from Eq.~(\ref{eq:negf}) and Eq.~(\ref{eq:time}) to prove Eq.~(\ref{eq:ref}). 

The spin-transfer-torque conductances are not related to one another with universal reciprocity relations, because the currents and occupations functions are defined at different cross-sections. 

\section{General Circuit Representation}
\label{sec:circuit}

\begin{figure}
\includegraphics[width=\linewidth]{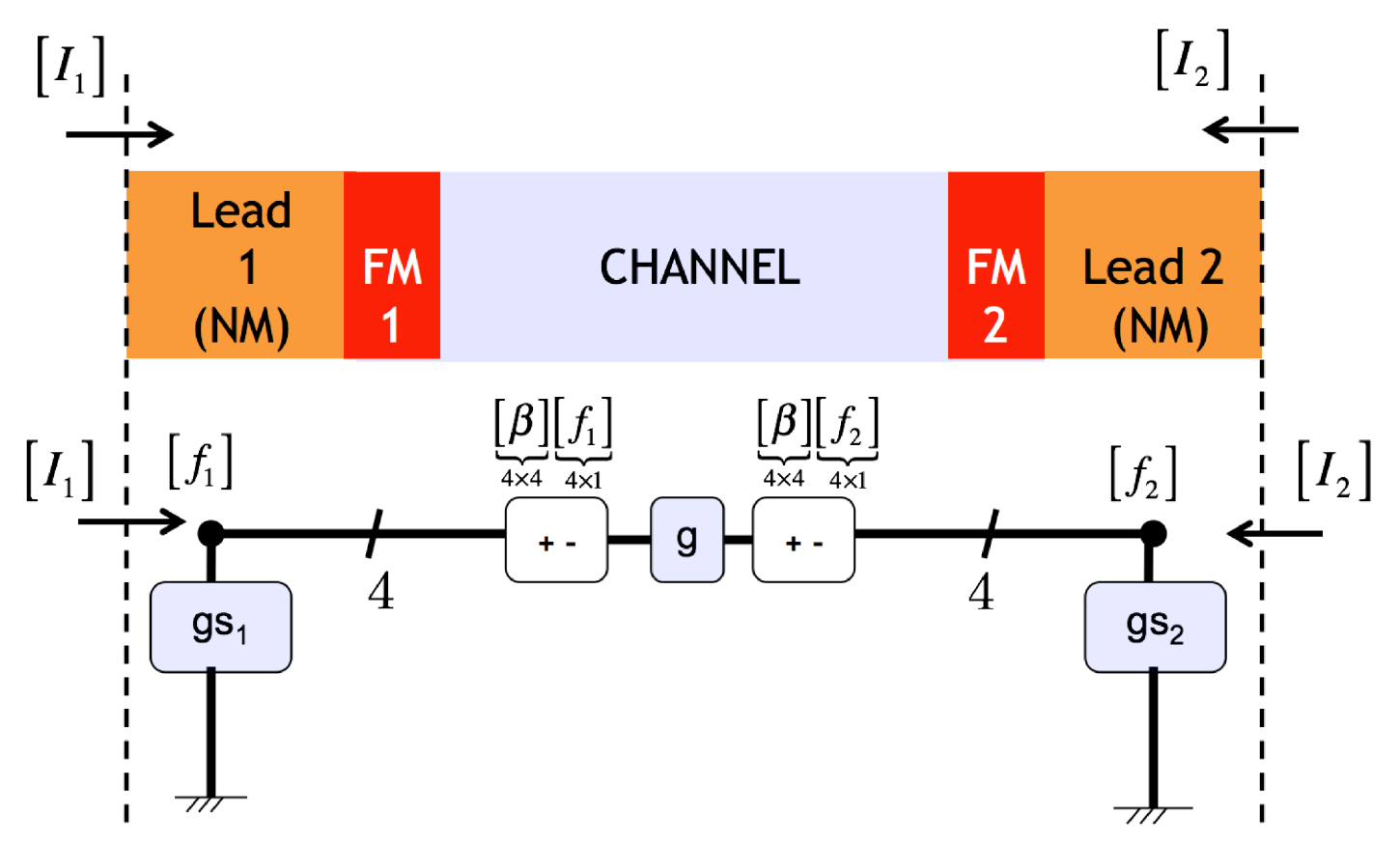}
\caption{Four component circuit representation of a 2-Terminal spin device with an arbitrary channel, and magnetic contacts. All nodes carry 4 currents and 4 voltages, while all circuit elements are $4\times4$ matrices, uniquely defined in terms of conductances shown in Eq.~(\ref{eq:negf}).}
\label{fi:cir}
\end{figure}

Fig.~(\ref{fi:cir}) shows a possible circuit representation of the terminal conductance matrices that can be readily implemented in  SPICE-like circuit simulators.  The example is a 2-Terminal structure for simplicity; however, a similar circuit can be implemented for a conductor with any number of terminals.  The circuit components that are shown in FIG.~(\ref{fi:cir}) are all uniquely defined in terms of the terminal conductances:
\begin{equation}\begin{gathered}
  g{s_1} = {\widetilde{G}_{11}} + {\widetilde{G}_{21}}\quad {\text{         }}g{s_2} = {\widetilde{G}_{22}} + {\widetilde{G}_{12}} \hfill \\
  g =  - ({\widetilde{G}_{12}} + {\widetilde{G}_{21}})/2\quad{\text{        }}\beta ={g^{ - 1}}({\widetilde{G}_{21}} - {\widetilde{G}_{12}})/2 \hfill \\ 
\end{gathered}
\label{eq:cir} 
\end{equation}
making each of these conductances are $4\times 4$ matrices.

Note that the circuit elements defined in Eq.~(\ref{eq:cir}) are generic and can be used with 4-component conductances based on other microscopic theories, such as Scattering Theory. 
%There are two different features of the circuit shown in FIG.~(\ref{fi:cir}) compared to ordinary charge-based circuits: a) Shunt conductances, a consequence of recombination (generation) of spin-currents within the conductor and b) Voltage Controlled Voltage Sources (VCVS), a consequence of a possible non-reciprocity of the general network.   
The shunt conductances $gs_1$ and $gs_2$ account for differences in spin currents entering and leaving the device whose $cc$ and $cs$  elements are zero, prohibiting charge currents through these conductances. The Voltage Controlled Voltage Sources (VCVS) are required since  a physically asymmetric device having a +z magnet on the left and a +x magnet on the right, will ``look'' different to an incoming spin current from both sides. Similar VCVS elements have been used to describe non-reciprocal circuits of the classical Hall Effect for possible circuit implementations\cite{ramsden_hall-effect_2011}. 

For self-consistent magnetization dynamics and transport simulations, the same circuit implementation can be used with spin-transfer-torque conductances that would be supplied to an LLG solver in a SPICE implementation.

\section{From Hamiltonians To Circuit Models: A Few Examples}
\label{sec:hamtocirc}
In general our main result Eq.~\ref{eq:negf} can be applied once the Green's function for the channel has been calculated, either analytically or numerically for any geometry in a model-independent manner, with the assumption that the transport can be treated in the elastic, coherent regime. After this step, the obtained conductances can easily be represented as the spin-circuit  described in Fig.~\ref{fi:cir} and then efficiently simulated in a SPICE-like simulator like any other charge-based conductor. 

In this section we apply Eq.~(\ref{eq:negf}) to two model Hamiltonians describing Rashba and Hanle effects respectively. In accordance with Eq.~(\ref{eq:negf}), our methodology assumes coherent and ballistic transport, furthermore in the analysis below, we will first assume that the transport is in 1D and then perform a mode summation in real space assuming periodic boundary conditions in the y-direction, as commonly done in device analysis within the NEGF framework \cite{zainuddin2011voltage}.

\subsection{Rashba Spin-Orbit (RSO) Coupling Materials: 1D}
The  Hamiltonian that describes the Rashba interfaces is given by:
\begin{equation}
H = H_0 + \eta(\sigma_x k_y - \sigma_y k_x) 
\end{equation}
where $\sigma$ denotes the Pauli spin matrices and $H_0$ is the Hamiltonian of the unperturbed system and $\eta$ is the Rashba coefficient in units of $J-m$. Starting from the fundamental mode for transport, $k_y=0$, we show that (Supplementary Information) the $4\times 4$ conductances describing the system can be written as:
\begin{equation}
\label{eq:mat}
\begin{pmatrix}
I_1 \\
I_2 
\end{pmatrix}=\begin{bmatrix}
G_{11} & G_{12} \\
G_{21} & G_{22}
\end{bmatrix}\begin{pmatrix}
V_1 \\
V_2 
\end{pmatrix}
\end{equation}
where the  conductance matrices (per spin) in 1D ($k_y=0$) for a Rashba Spin-Orbit (RSO) channel with NM leads  can be obtained using Eq.~(\ref{eq:negf}):
\begin{equation}
G_{11}=G_{22}= \left(\frac{q^2 \ }{h}\right) 
\begin{bmatrix}  & c & z & x & y \\
              c & 1 & 0 & 0 & 0 \\
              z & 0 & 1  & 0 & 0 \\
              x & 0 & 0 & 1 & 0  \\
              y & 0 & 0 & 0 & 1
							\end{bmatrix}
\end{equation} 

$G_{11}$ and $G_{22}$ are simply the interface conductances due to the ideal NM contacts. $G_{12}$ and $G_{21}$ read:

\begin{equation}
G_{12}= -\left(\frac{q^2 \ }{h}\right) 
\begin{bmatrix}  & c & z & x & y \\
              c & 1 & 0 & 0 & 0 \\
              z & 0 & \cos(\theta)  & \sin(\theta) & 0 \\
              x & 0 & -\sin(\theta) & \cos(\theta) & 0  \\
              y & 0 & 0 & 0 & 1
							\end{bmatrix} \\
\end{equation}
\begin{equation}
G_{21}= -\left(\frac{ q^2 \ }{h}\right) 
\begin{bmatrix}  & c & z & x & y \\
              c & 1 & 0 & 0 & 0 \\
              z & 0 & \cos(\theta)  & -\sin(\theta) & 0 \\
              x & 0 & \sin(\theta) & \cos(\theta) & 0  \\
              y & 0 & 0 & 0 & 1
\end{bmatrix}
              \label{eq:rashba1D}
\end{equation}
Note that the reciprocity relation is satisfied according to Eq.~(\ref{eq:ref}). The rotation angle is given by:
\begin{equation}
\theta=\frac{2m^*\eta L}{  \hbar^2}
\label{eq:rso_rot}
\end{equation}
where $m^*$ and  $L$ are effective mass and length of the channel and $\eta$ is the Rashba coefficent and $\hbar$ is the reduced Planck's constant, in an effective mass approximation \cite{zainuddin2011voltage}.  The 1D results obtained for the $G_{12}$ and $G_{21}$ conductances are intuitively appealing, since the rotation angle can be related to the product of the angular velocity due to the splitting of the bands due to the `effective' y-directed magnetic field felt by the electrons due to the Rashba interaction and the time spent in the channel since  $\omega = 2\eta k_x/\hbar$ and $t = L m^* /\hbar k_x$ making $\theta=\omega t$.

\subsection{Rashba Spin-Orbit (RSO) Coupling Materials: 2D}

So far our analysis has been in 1D. However, in a 2D conductor the transport is not limited to the fundamental mode $k_y=0$. Assuming periodic boundary boundary conditions in the transverse direction, the effect of higher order modes can be summed to obtain an average 2D conductance. In the case of Rashba Effect, the angular spread gives rise to two different effects: (a) The effective field is momentum dependent, making the rotation axis different for each mode (b) The time-of-flight of electrons are different, changing the total time spent in the magnetic field. The second effect can be  included to the rotation angle, $\theta$ through the length of the channel, $L_{eff}= L/\cos(\phi)$ where $\phi  = \tan^{-1}(k_y/k_x) $, below $\theta'=\theta/\cos\phi$. The first effect can be included by rotating the 1D result around the z-axis by angle $\phi$ to obtain the conductance matrices in an arbitrary $(k_x,k_y)$ combination as follows:
\begin{strip}
\begin{equation}
%G_{12}(\phi)= \left[ \begin {array}{cccc} 1&0&0&0\\ \noalign{\medskip}0&\cos
 %\left( \theta' \right) &\sin \left( \theta' \right) \cos \left( \phi
 %\right) &-\sin \left( \theta' \right) \sin \left( \phi \right) 
%\\ \noalign{\medskip}0&-\sin \left( \theta' \right) \cos \left( \phi
% \right) &\cos \left( \theta' \right)  \left( \cos \left( \phi \right) 
% \right) ^{2}+1- \left( \cos \left( \phi \right)  \right) ^{2}&-\sin
% \left( \phi \right) \cos \left( \phi \right)  \left( \cos \left( 
%\theta' \right) -1 \right) \\ \noalign{\medskip}0&\sin \left( \theta'
% \right) \sin \left( \phi \right) &-\sin \left( \phi \right) \cos
% \left( \phi \right)  \left( \cos \left( \theta' \right) -1 \right) &
%\cos \left( \theta' \right) -\cos \left( \theta' \right)  \left( \cos
% \left( \phi \right)  \right) ^{2}+ \left( \cos \left( \phi \right) 
% \right) ^{2}\end {array} \right] 
% \label{eq:rashba2D}
%\end{equation}
G_{12}(\phi)=\left[
\begin{array}{cccc}
 1 & 0 & 0 & 0 \\
 0 & \cos (\theta' ) & \cos (\phi ) \sin (\theta' ) & -\sin (\theta' ) \sin (\phi ) \\
 0 & -\cos (\phi ) \sin (\theta' ) & \cos (\theta' ) \cos ^2(\phi )+\sin ^2(\phi ) & -(\cos (\theta' )-1) \cos (\phi ) \sin (\phi ) \\
 0 & \sin (\theta' ) \sin (\phi ) & -(\cos (\theta' )-1) \cos (\phi ) \sin (\phi ) & \cos ^2(\phi )+\cos (\theta' ) \sin ^2(\phi ) \\
\end{array}
\right],
              \label{eq:rashba2D}
\end{equation}
\end{strip}
\noindent To obtain an average 2D conductance from Eq.~\ref{eq:rashba2D}, we need to sum over these modes and average them. We define $s = k_y / k_f $ so that $\cos\phi=\sqrt{1-s^2}$, making $\theta'=\theta_0/\sqrt{1-s^2}$. The average conductance can be expressed as:
\begin{equation}
G_{12}^{(2D)}  = \displaystyle{\displaystyle\int_{s=0}^{s=1} ds \ G_{12(s)}}
\label{eq:save}
\end{equation}

Eq.~\ref{eq:save} can be evaluated exactly using the stationary phase approximation \cite{bleistein1975asymptotic} (Supplementary Information) yielding an analytical conductance matrix for the Rashba Hamiltonian in 2D. This result fully captures the spin-relaxation due to the effect of mode mixing ($L_{eff}=L/\sqrt{1-s^2})$ and the effect of a momentum dependent magnetic field, $G_{12}^{(2D)}$ is given as:

\footnotesize
\begin{equation}
%G_{12}^{2D}= \left[ \begin {array}{cccc} 1&0&0&0\\ \noalign{\medskip}0&\sqrt{\displaystyle\frac{\pi}{2\theta}}\cos
% \left( \theta+\pi/4\right) & \sqrt{\displaystyle\frac{\pi}{2\theta}}\sin
% \left( \theta+\pi/4\right) & 0 
%\\ \noalign{\medskip}0& -\sqrt{\displaystyle\frac{\pi}{2\theta}}\sin
% \left( \theta+\pi/4\right)  & \displaystyle\frac{1}{3}+\sqrt{\displaystyle\frac{\pi}{2\theta}}\cos
% \left( \theta+\pi/4\right)  &0 \\ \noalign{\medskip}0&0  &0 &\displaystyle \frac{2}{3} \end {array} \right] 
\hspace{-5pt} \left[ \begin {array}{cccc} 1&0&0&0\\ \noalign{}0&\sqrt{\displaystyle\frac{\pi}{2\theta}}\cos
 \left( \theta+\pi/4\right) & \sqrt{\displaystyle\frac{\pi}{2\theta}}\sin
 \left( \theta+\pi/4\right) & 0 
\\ \noalign{}0& -\sqrt{\displaystyle\frac{\pi}{2\theta}}\sin
 \left( \theta+\pi/4\right)  & \displaystyle\frac{1}{3}+\sqrt{\displaystyle\frac{\pi}{2\theta}}\cos
 \left( \theta+\pi/4\right)  &0 \\ \noalign{}0&0  &0 &\displaystyle \frac{2}{3} \end {array} \right] 
 \label{eq:rashba2D_spa} 
 \end{equation}
%\end{strip}
\normalsize

 where we have normalized the conductance with the ballistic conductance (per spin) $G_B = \displaystyle (q^2/h)\  M$. We point out that even when the transport is not entirely ballistic, the prefactor can be adjusted based on experimental values of ordinary conductance while the off-diagonal elements capture the essential precession physics as observed in experiments demonstrating the Rashba effect \cite{koo2009control,choi2015electrical}. Moreover, the ballistic transport assumption can easily be relaxed through including impurity scattering to the Hamiltonian in the NEGF formalism and  numerical  conductances can still be obtained from Eq.~(\ref{eq:negf}), however obtaining simple analytical formulations as shown in \ref{eq:rashba2D_spa} may not be possible.  In fig.~(\ref{fi:fig3}) we compare various entries of this analytical  matrix with the exact numerical integration and find that they are in excellent agreement. We have numerically tested all other entries of the matrix and found that they agree as well. We also note that the diagonal entries presented here are also in agreement with the NEGF-based analysis shown in \cite{zainuddin2011voltage}. The 2D result given in  Eq.~(\ref{eq:rashba2D_spa}) can be used in device models involving Rashba spin-orbit materials, as a modular building block\cite{camsari2015modular}.

 \subsection{Hanle Conductances in 2D}

 Following a very similar analysis for the Hanle Hamiltonian: 
 
\begin{equation}
 H= H_0 + \mu_B \sigma_z B_z 
 \end{equation}

We first obtain 1D conductances in presence of a z-directed magnetic field, and then do an analytical mode summation to obtain 2D conductance, $G_{12}^{(2D)}$:

\footnotesize
\begin{equation}
 \left[ \begin {array}{cccc} 1&0&0&0\\ \noalign{}0&1 & 0 & 0 
\\ \noalign{}0& 0  & \sqrt{\displaystyle\frac{\pi}{2| \theta| }}\cos
 \left( \mathrm{sgn}(\theta)+\pi/4\right)  &\sqrt{\displaystyle\frac{\pi}{2| \theta| }}\sin
 \left( \mathrm{sgn}(\theta)+\pi/4\right)  \\ \noalign{}0&0  &-\sqrt{\displaystyle\frac{\pi}{2| \theta| }}\sin
 \left( \mathrm{sgn}(\theta)+\pi/4\right)  &\sqrt{\displaystyle\frac{\pi}{2| \theta| }}\cos
 \left( \mathrm{sgn}(\theta)+\pi/4\right)  \end {array} \right] 
 \label{eq:Hanle2D_spa}
\end{equation}
\normalsize

where the rotation angle is given as  $\theta= 2\mu_B B_z/ \hbar  L m^*/\hbar |k_f|$ which also be expressed in terms of the angular velocity $\omega=2\mu_B B/\hbar$, which is not momentum-dependent,  and time of flight, $\tau=Lm^*/\hbar |k_f|$. Unlike the Rashba effect, the magnetic field can be varied between negative and positive values which is reflected in the final result. FIG.~\ref{fi:fig3} shows the diagonal entries of this conductance matrix along with the exact mode summation. The physics of 2D spin-precession captured in this conductance can be used in the analysis of recent experiments that combine the physics of spin-Hall effect in doped graphene with the Hanle effect \cite{balakrishnan2013colossal}.  Even though we have assumed the external magnetic field in the +z direction, Eq.~(\ref{eq:Hanle2D_spa}) can be transformed to any arbitrary direction using an appropriate rotation matrix. 

\begin{figure}[tpb]
\includegraphics[width=\linewidth]{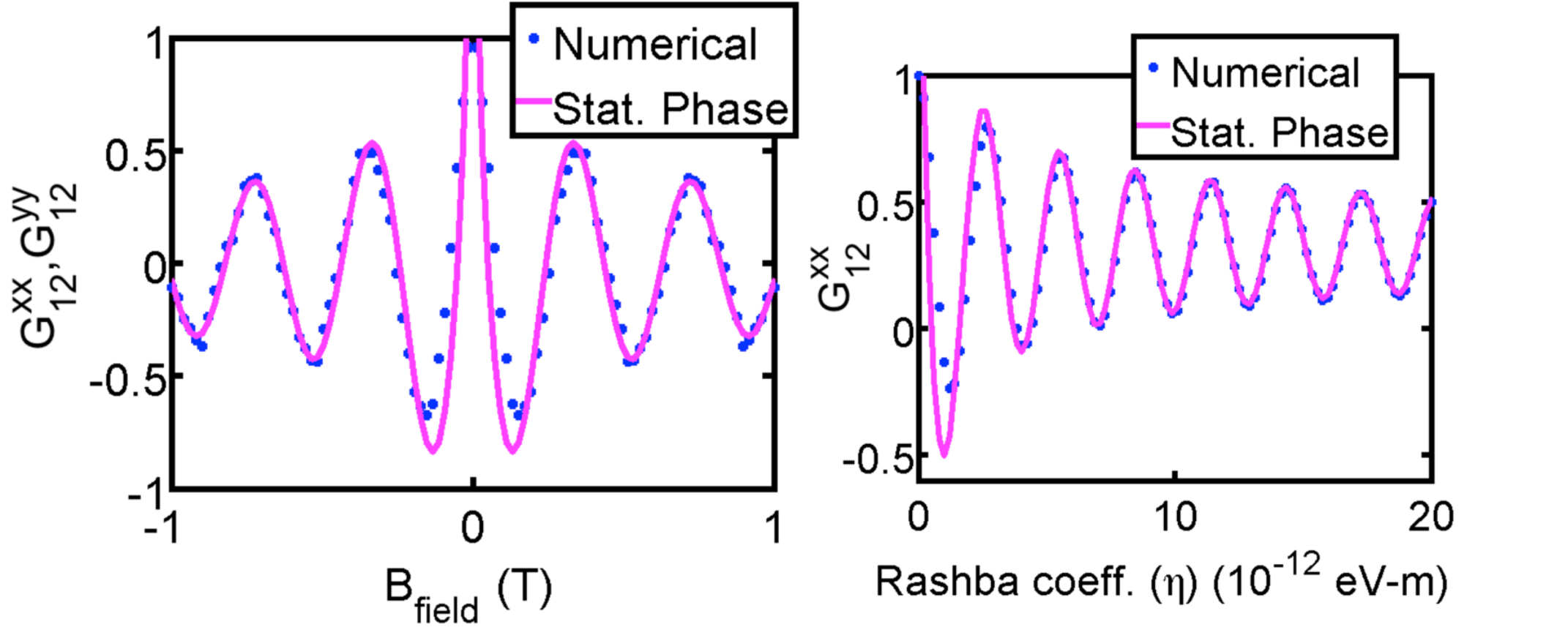}
\caption{Selected conductance elements for Rashba and Hanle effect as a function of external magnetic field (z-directed) and the Rashba coefficient. The analytical 2D conductances obtained using the stationary phase approximation are compared with an exact summation of modes showing close agreement. The 1D conductances that are obtained from Eq.~(\ref{eq:negf}) are numerically checked against direct NEGF calculations. The conductances are normalized with the ballistic conductance (per spin), $(q^2/h) M $. The numerical parameters used to calculate the rotation angle and the integrated conductances in this simulation are: $k_f  = 4.118\times 10^8 \ \rm m^{-1}$, $L=1.65 \  \mu \rm m$ and $m^*=0.05 \ m_0$ and $m^*=0.5 \ m_0$ for Rashba and Hanle effects respectively. }
\label{fi:fig3}
\end{figure} 

\section{Conclusions and summary of main results}
Existing NEGF-based models for charge-based nanoelectronic devices often use the conductance expression,  $ \rm tr. \left[\Gamma_1 G^R \Gamma_2 G^A \right]$ in the elastic, coherent transport regime. Recent advances in spintronics have raised the possibility of spin potential-driven electronic devices. In this paper:
\begin{itemize}
\item We provided conductance expressions involving spin potentials and spin currents generalizing the charge conductance $G^{cc}$ of NEGF-based model\cite{Meir_Wingreen}.
\item We provided  ``spin-transfer-torque" conductances to be used in self-consistent simulations of magnetization dynamics and spin transport.
\item We have shown that these conductances pass critical tests by automatically  satisfying universal sum rules and reciprocity relations that  must hold irrespective of the details of transport.
\item We have represented our results in a generic 4-component circuit to be implemented in SPICE-like simulators towards a unified description of hybrid devices involving existing CMOS and spintronic circuit elements.

\item We have  applied this formalism (using Eq.~\ref{eq:negf}) to obtain analytical conductances that describe the Hanle and the Rashba channels in 2D.
\end{itemize} 

Finally, even though our analysis here was restricted to the spin degree of freedom, our general methodology can be applied to obtain generalized conductances that would capture other degrees of freedom, such as valley and pseudospin\cite{xu2014spin}  that would bridge  their description from quantum transport to ordinary circuits. 
\vspace{25pt}
\section*{Acknowledgment}
{This work was supported in part by C-SPIN, one of six centers of STARnet, a Semiconductor Research Corporation program, sponsored by MARCO and DARPA and in part by the National Science Foundation through the NCN-NEEDS program, contract 1227020-EEC.}

\noindent
\bibliographystyle{ieeetr}
\balance\bibliography{bibNEGF2015}
%\vspace*{3pt}


\section*{Supplementary material}

# Supplementary Information: Non-Equilibrium Green's Function based Circuit Models for Coherent Spin Devices

Kerem Yunus Camsari,[1, *] Samiran Ganguly,[2] Deepanjan Datta,[3] and Supriyo Datta[1]

[1]*School of Electrical and Computer Engineering, Purdue University, IN, 47907*
[2]*Dept. of Electrical and Computer Engineering, University of Virginia, VA, 22904*
[3]*GLOBALFOUNDRIES, Bangalore, India*

## I. SUM RULES

In this section, we prove Eq. (14,15) starting from Eq. (10).

$$\sum_{m=1}^{P} \left[ \widetilde{G}_{mn} \right]^{cc} = 0$$

Starting from Eq. (10) for $cc$ elements:

$$\sum_{m} \left[ \widetilde{G}_{mn} \right]^{cc} = \frac{q^2}{h} \sum_{m} \left( \begin{matrix} i \ \text{tr.} \left[ \Gamma_m G^R - \Gamma_m G^A \right] \delta_{mn} \\ -\text{tr.} \left[ \Gamma_m G^R \Gamma_n G^A \right] \end{matrix} \right)$$

And using the following NEGF identities:

$$A = i \left[ G^R - G^A \right] = \sum_{k} G^R \Gamma_k G^A = \sum_{k} G^A \Gamma_k G^R$$

The desired sum rule then, follows:

$$\sum_{m} \left[ \widetilde{G}_{mn} \right]^{cc} = \frac{q^2}{h} \left( \begin{matrix} \text{tr.} \ \left[ \Gamma_n A \right] \\ -\text{tr.} \left[ \underbrace{G^A \sum_{m} \Gamma_m G^R}_{A} \Gamma_n \right] \end{matrix} \right) = 0$$

$$\sum_{m=1}^{P} \left[ \widetilde{G}_{mn} \right]^{cs} = 0$$

Specializing to $cs$ elements (s representing spin indices x,y,z):

$$\sum_{m} \left[ \widetilde{G}_{mn} \right]^{cs} = \frac{q^2}{h} \sum_{n} \left( \begin{matrix} i \ \text{tr.} \left[ S_\beta \Gamma_m G^R - S_\beta \Gamma_m G^A \right] \delta_{mn} \\ -\text{tr.} \ \left[ \Gamma_m G^R S_\beta \Gamma_n G^A \right] \end{matrix} \right)$$

using the same NEGF identities, we obtain:

$$\sum_{m} \left[ \widetilde{G}_{mn} \right]^{cs} = \frac{q^2}{h} \left( \begin{matrix} \text{tr.} \ \left[ S_\beta \Gamma_n A \right] \\ -\text{tr.} \left[ \underbrace{G^A \sum_{m} \Gamma_m G^R}_{A} S_\beta \Gamma_n \right] \end{matrix} \right) = 0$$

* kcamsari@purdue.edu

$$\sum_{n=1}^{P} \left[ \widetilde{G}_{mn} \right]^{cc} = 0$$

Starting from Eq. (10) for cc:

$$\sum_{n} \left[ \widetilde{G}_{mn} \right]^{cc} = \frac{q^2}{h} \sum_{n} \left( \begin{array}{l} i \text{ tr. } \left[ \Gamma_m G^R - \Gamma_m G^A \right] \delta_{mn} \\ -\text{tr. } \left[ \Gamma_m G^R \Gamma_n G^A \right] \end{array} \right)$$

Making use of the NEGF identities shown above:

$$\sum_{n} \left[ \widetilde{G}_{mn} \right]^{cc} = \frac{q^2}{h} \left( \begin{array}{l} \text{tr. } \left[ \Gamma_m A \right] \\ \\ -\text{tr. } \left[ \Gamma_m \ G^R \underbrace{\sum_{n} \Gamma_n G^A}_{A} \right] \end{array} \right) = 0$$

$$\sum_{n=1}^{P} \left[ \widetilde{G}_{mn} \right]^{sc} = 0$$

Specializing to $sc$ elements for the conductance matrices:

$$\sum_{n} \left[ \widetilde{G}_{mn} \right]^{sc} = \frac{q^2}{h} \sum_{n} \left( \begin{array}{l} i \text{ tr. } \left[ S_\alpha \Gamma_m G^R - S_\alpha \Gamma_m G^A \right] \delta_{mn} \\ -\text{tr. } \left[ S_\alpha \Gamma_m G^R \Gamma_n G^A \right] \end{array} \right)$$

we have:

$$\sum_{n} \left[ \widetilde{G}_{mn} \right]^{sc} = \frac{q^2}{h} \left( \begin{array}{l} \text{tr.} \left[ S_\alpha \Gamma_m A \right] - \\ \\ \text{tr.} \left[ S_\alpha \Gamma_m \underbrace{\sum_{n} G^R \Gamma_n G^A}_{A} \right] \end{array} \right) = 0$$

## II.  RECIPROCITY

In this section, we start from Eq. (10) to prove Eq. (17)

$$\left[ \widetilde{G}_{mn} \right]^{\alpha\beta} \Big|_{+B} = \frac{q^2}{h} \left( \begin{array}{l} \text{tr. } \left[ i S_\beta S_\alpha \Gamma_m G^R - S_\alpha S_\beta \Gamma_m G^A \right] \delta_{mn} \\ -\text{tr. } \left[ S_\alpha \Gamma_m G^R S_\beta \Gamma_n G^A \right] \end{array} \right)_{+B}$$

and write the equivalent expression for $B$ using the time-reversibility rules for Green's functions (Eq. (18) in the text):

$$= \frac{q^2}{h} \left( \begin{array}{l} -\text{tr. } \left[ S_\alpha (\Gamma_m)^T \left( S_y (G^R)^T S_y \right) S_\beta (\Gamma_n)^T \left( S_y (G^A)^T S_y \right) \right] + \\ \\ i \text{ tr. } \left[ \begin{array}{l} S_\beta S_\alpha (\Gamma_m)^T \left( S_y (G^R)^T S_y \right) - \\ S_\alpha S_\beta (\Gamma_m)^T \left( S_y (G^A)^T S_y \right) \end{array} \right] \delta_{mn} \end{array} \right)_{-B}$$

Note that the broadening matrices $\Gamma$ only pick up a transpose under time reversal because they correspond to non-magnetic contacts with no spin preference. First we observe:

$$S_y S_\alpha = (S_\alpha)^T S_y (-1)^{n_\alpha}$$

where $\alpha$ is a spin or charge index, and $n_\alpha$ is 1 for spin indices and 0 for charge. Next, using the cyclic property of traces with the fact that $\Gamma$ commutes with Pauli matrices (non-magnetic contacts):

$$= \frac{q^2}{h} \left( \begin{array}{l} -\mathrm{tr.} \left[ \underbrace{S_y S_\alpha S_y}_{S_\alpha{}^T (-1)^{n_\alpha}} \, (\Gamma_m)^T (G^R)^T \, \underbrace{S_y S_\beta S_y}_{S_\beta{}^T (-1)^{n_\beta}} \, (\Gamma_n)^T (G^A)^T \right] + \\[3em] i \, \mathrm{tr.} \left[ \begin{array}{l} \underbrace{S_y S_\beta S_\alpha S_y}_{S_\beta{}^T S_\alpha{}^T (-1)^{(n_\alpha + n_\beta)}} \, (\Gamma_m)^T (G^R)^T - \\[2em] \underbrace{S_y S_\alpha S_\beta S_y}_{S_\alpha{}^T S_\beta{}^T (-1)^{(n_\alpha + n_\beta)}} \, (\Gamma_m)^T (G^A)^T \end{array} \right] \delta_{mn} \end{array} \right)_{-B}$$

simplifying to:

$$= \frac{q^2}{h} \left( \begin{array}{l} -\mathrm{tr.} \left[ \left( S_\beta \Gamma_n G^R S_\alpha \Gamma_m G^A \right)^T \right] (-1)^{n_\alpha + n_\beta} + \\[1em] i \, \mathrm{tr.} \left[ \left( G^R \Gamma_m S_\alpha S_\beta - G^A \Gamma_m S_\beta S_\alpha \right)^T \right] \delta_{mn} (-1)^{(n_\alpha + n_\beta)} \end{array} \right)_{-B}$$

the right hand side is by definition defined as:

$$[G_{mn}]^{\alpha\beta} \Big|_{+B} = [G_{nm}]^{\beta\alpha} \Big|_{-B} (-1)^{n_\alpha + n_\beta}$$

since trace of $A^T$ is equal to trace A, proving the reciprocity relation shown in the main text.

## III. RECOMBINATION-GENERATION

In this section, we start from Eq. (5) and show that it can be decomposed into the sum of two distinct components, Eq. (7) and Eq. (8). Starting from:

$$\mathrm{tr.} \left[ S_\alpha I_{op} \right] = \frac{q}{h} \mathrm{tr.} \left[ i S_\alpha \begin{pmatrix} G^n H - H G^n + \\ G^n \Sigma^\dagger - \Sigma \, G^n + \\ G^R \Sigma^{in} - \Sigma^{in} G^A \end{pmatrix} \right] \tag{1}$$

Using the cyclic property of trace operator:

$$\mathrm{tr.} \left[ i \left( H S_\alpha - S_\alpha H \right) G^n \right] +$$
$$\mathrm{tr.} \left[ i \left( \Sigma^\dagger S_\alpha - S_\alpha \Sigma \right) G^n \right] +$$
$$\mathrm{tr.} \left[ i \left( S_\alpha G^R - G^A S_\alpha \right) \Sigma^{in} \right]$$

In explicit notation where $(m, n)$ refer to lattice sites where each $A_{mn}$ below corresponds to a $2 \times 2$ spin matrix:

$$\sum_{n,m} \left[ i (H S_\alpha - S_\alpha H)_{n,m} (G^n)_{m,n} \right] +$$
$$\sum_{n,m} \left[ i \left( \Sigma^\dagger S_\alpha - S_\alpha \Sigma \right)_{n,m} (G^n)_{m,n} \right] +$$
$$\sum_{n,m} \left[ i \left( S_\alpha G^R - G^A S_\alpha \right)_{n,m} \left( \Sigma^{in} \right)_{m,n} \right]$$

**First Term:** The first coefficient in the first term reads:

$$i(HS_\alpha - S_\alpha H)_{n,m}$$
$$= \sum_k H_{n,k}(S_\alpha)_{k,m} - (S_\alpha)_{n,k}H_{k,m}$$
$$= H_{n,m}(S_\alpha)_{m,m} - (S_\alpha)_{n,n}(H)_{n,m}$$
$$= H_{n,m}\sigma - \sigma H_{n,m}$$

Inside the non-magnetic contacts, $H_{n,m}$ commutes with all Pauli spin matrices for $(m,n) \in$ contact boundaries, making this term identically zero along all contact regions.

**Second Term:** The first coefficient in the second term of reads:

$$\left(\Sigma^\dagger S_\alpha - S_\alpha \Sigma\right)_{n,m}$$
$$= \sum_k \Sigma^\dagger{}_{n,k}(S_\alpha)_{k,m} - (S_\alpha)_{n,k}\Sigma_{k,m}$$
$$= \Sigma^\dagger{}_{n,m}(S_\alpha)_{m,m} - (S_\alpha)_{n,n}\Sigma_{n,m}$$
$$= \Sigma^\dagger{}_{n,m}\sigma - \sigma\ \Sigma_{n,m}$$

The term $\Sigma^\dagger_{n,m}$ is zero unless $(m,n)$ are lattice points within contact boundaries, making the second term non-zero, only along contact regions.

**Third Term:** The third term will be zero unless $\Sigma^{in}_{m,n}$ is non-zero. By definition:

$$\Sigma^{in} = \gamma \otimes [f]_{2\times2}$$

where $\gamma$ is the $N \times N$ broadening matrix, $N$ being the number of lattice points while $\gamma_{m,n}$ is non-zero only for $(m,n)$ that are lattice points within contact boundaries. In summary, second and third term can be associated with Eq. (7) and the first term with Eq. (8).

## IV. RASHBA AND HANLE CONDUCTANCES

In this section we show an analytical justification for the main paper equations, Eq. (24-26) directly from Eq. (10), the analysis for Eq. (30) is very similar and therefore omitted. Note that Eq. (24-26) have been extensively verified with numerical examples, for much larger channel lengths compared to what we analytically investigate here. We start from a tight-binding Hamiltonian that describes Rashba interfaces:

$$H = H_0 + \alpha(\sigma_x k_y - \sigma_y k_x) \tag{2}$$

where $\alpha$ is the Rashba coefficient, in units $J - m$.

We start from a tight-binding Hamiltonian for a channel that has only 2 discrete points making all matrices $4 \times 4$, as a minimum Hamiltonian to exhibit spin-precession. Following the tight-binding prescription outlined in Chapter 22 of [1] for Rashba Hamiltonians:

$$H = \begin{bmatrix} 2\,t & 0 & t & 1/2\,\eta \\ 0 & 2\,t & -1/2\,\eta & t \\ t & -1/2\,\eta & 2\,t & 0 \\ 1/2\,\eta & t & 0 & 2\,t \end{bmatrix} \tag{3}$$

where we have defined $\eta = \alpha/a$, where $a$ is the lattice spacing so that $\eta$ is in units of energy.

The NM leads are desribed by $\Sigma$ matrices:

$$\Sigma_L = \begin{bmatrix} -te^{ika} & 0 & 0 & 0 \\ 0 & -te^{ika} & 0 & 0 \\ 0 & 0 & 0 & 0 \\ 0 & 0 & 0 & 0 \end{bmatrix} \quad \Sigma_R = \begin{bmatrix} 0 & 0 & 0 & 0 \\ 0 & 0 & 0 & 0 \\ 0 & 0 & -te^{ika} & 0 \\ 0 & 0 & 0 & -te^{ika} \end{bmatrix} \tag{4}$$

Assuming a dispersion of $E = 2t \left(1 - cos(ka)\right)$ and from the definition of the retarded Green's function, $G^R = [EI - H - \Sigma_L - \Sigma_R]^{-1}$, we get $(G^R)^{-1}$ as :

$$\begin{bmatrix} -t\cos\left(x\right) + it\sin\left(x\right) & 0 & t & 1/2\,\eta \\ 0 & -t\cos\left(x\right) + it\sin\left(x\right) & -1/2\,\eta & t \\ t & -1/2\,\eta & -t\cos\left(x\right) + it\sin\left(x\right) & 0 \\ 1/2\,\eta & t & 0 & -t\cos\left(x\right) + it\sin\left(x\right) \end{bmatrix}$$

We take the inverse analytically and group the result in terms of $2 \times 2$ matrices,

$$G^R \equiv \begin{bmatrix} g_{11}^R & g_{12}^R \\ g_{21}^R & g_{22}^R \end{bmatrix} \tag{5}$$

Using these matrices in Eq. (10) to obtain the conductances directly, we have for $G_{21}$:

$$\underbrace{[G_{21}]^{\alpha\beta}}_{4\times4\ \text{spin-conductances}} = \frac{1}{2}\text{Tr.}[\sigma_\alpha\ \gamma_{22}\ g_{21}^R\ \gamma_{11}\ \sigma_\beta\ g_{12}^A] \tag{6}$$

where $g^A = (g^R)^\dagger$ and $\sigma_{\alpha,\beta}$ are Pauli spin matrices for different spin orientations and $\gamma_{i,j}$ are $2 \times 2$ broadening matrices[1]. Taking the full inverse in Eq. (5), we calculate $g_{21}^R$ to be:

$$g_{21}^R = \begin{bmatrix} 4\,\dfrac{(i\sin\left(x\right) + \cos\left(x\right))\,t}{8\,it^2\sin\left(x\right) + i\eta^2\sin\left(x\right) + \eta^2\cos\left(x\right)} & -2\,\dfrac{(i\sin\left(x\right) + \cos\left(x\right))\,\eta}{8\,it^2\sin\left(x\right) + i\eta^2\sin\left(x\right) + \eta^2\cos\left(x\right)} \\ 2\,\dfrac{(i\sin\left(x\right) + \cos\left(x\right))\,\eta}{8\,it^2\sin\left(x\right) + i\eta^2\sin\left(x\right) + \eta^2\cos\left(x\right)} & 4\,\dfrac{(i\sin\left(x\right) + \cos\left(x\right))\,t}{8\,it^2\sin\left(x\right) + i\eta^2\sin\left(x\right) + \eta^2\cos\left(x\right)} \end{bmatrix} \tag{7}$$

and the broadening matrices can be written from $\Sigma$ matrices as $\Gamma = i(\Sigma - \Sigma^\dagger)$:

$$\gamma_{11} = \gamma_{22} = \begin{bmatrix} 2\,t\sin\left(x\right) & 0 \\ 0 & 2\,t\sin\left(x\right) \end{bmatrix} \tag{8}$$

Using $g_{21}^R$ and $\gamma$ matrices in Eq. (6), we obtain $[G_{21}]^{zz}$ as:

$$[G_{21}]^{zz} = 16\,\frac{t^2\left(-4\,t^2 + \eta^2\right)\left(\sin\left(x\right)\right)^2}{64\left(\cos\left(x\right)\right)^2 t^4 - 64\,t^4 - \eta^4 + 16\left(\cos\left(x\right)\right)^2 t^2\eta^2 - 16\,t^2\eta^2} \tag{9}$$

Assuming a small perturbation due to the Rashba splitting, $\eta/t \ll 1$ the higher-order terms can be ignored. Defining $\theta = \eta/t$, in Eq. (9):

$$[G_{21}]^{zz} = \frac{4 - \theta^2}{4 + \theta^2} \tag{10}$$

Expanding this equation in series for small $\theta$:

$$[G_{21}]^{zz} = 1 - \frac{\theta^2}{2} + O(\theta^4) \tag{11}$$

which behaves like $\cos(\theta)$ for small $\theta$. Assuming $\sigma_\alpha = \sigma_z$ and $\sigma_\beta = \sigma_x$ in Eq. (6) we obtain:

$$G_{21}^{zx} = 64\,\frac{t^3\eta\left(\sin\left(x\right)\right)^2}{64\left(\cos\left(x\right)\right)^2 t^4 - 64\,t^4 - \eta^4 + 16\left(\cos\left(x\right)\right)^2 t^2\eta^2 - 16\,t^2\eta^2} \tag{12}$$

Defining $\theta = \eta/t$ and neglecting terms greater than $\theta^2$:

$$G_{21}^{zx} = -4\,\frac{\theta}{4 + \theta^2} \tag{13}$$

which can be expanded for small $\theta$:

$$G_{21}^{zx} = -\theta \tag{14}$$

which behaves like $-\sin(\theta)$ for small $\theta$. Earlier we set $\eta = \alpha(a)$ the rotation angle becomes,

$$\theta = \frac{\alpha}{ta} \tag{15}$$

Noting that t is originally defined as $t = \dfrac{\hbar^2}{2m^*a^2}$

$$\theta = \frac{2m^*\alpha(a)}{\hbar^2} \tag{16}$$

Setting $a = L$, we recover 1D-rotation angle for the channel length since in this example the rotation is only through a single lattice point $(\Delta L = a)$. The rest of the conductance matrices are obtained for different $\sigma_{\alpha,\beta}$ configurations similarly.

## V. 2D RSO CONDUCTANCES

The 1D conductances are rotated around the z-axis by angle $\phi$ to obtain the conductance matrices in an arbitrary $(k_x, k_y)$ combination which are under the influence of a mode-dependent effective magnetic fields:

$$G_{12}(\phi) = U_R G_{12}^{1D} U_R^\dagger \tag{17}$$

where

$$U_R = \begin{bmatrix} 1 & 0 & 0 & 0 \\ 0 & 1 & 0 & 0 \\ 0 & 0 & \cos(\phi) & \sin(\phi) \\ 0 & 0 & -\sin(\phi) & \cos(\phi) \end{bmatrix} \tag{18}$$

and $G_{12}^{1D}$ is:

$$\begin{bmatrix} 1 & 0 & 0 & 0 \\ 0 & \cos(\theta) & \sin(\theta) & 0 \\ 0 & -\sin(\theta) & \cos(\theta) & 0 \\ 0 & 0 & 0 & 1 \end{bmatrix} \tag{19}$$

Note that $\theta = \theta_0/\cos\phi$ for these modes. Substituting these in Eq. (17) we have:

$$\begin{bmatrix} 1 & 0 & 0 & 0 \\ 0 & \cos(\theta) & \sin(\theta)\cos(\phi) & -\sin(\theta)\sin(\phi) \\ 0 & -\sin(\theta)\cos(\phi) & \cos(\theta)(\cos(\phi))^2 + 1 - (\cos(\phi))^2 & -\sin(\phi)\cos(\phi)(\cos(\theta)-1) \\ 0 & \sin(\theta)\sin(\phi) & -\sin(\phi)\cos(\phi)(\cos(\theta)-1) & \cos(\theta)-\cos(\theta)(\cos(\phi))^2 + (\cos(\phi))^2 \end{bmatrix}$$

which is our stated result Eq. (28) in the main paper.

## VI. STATIONARY PHASE APPROXIMATION

In this section we show how we approximate integrals of the form:

$$\int_{s=0}^{s=1} ds \left( s^2 + s^2 \left( 1 - \cos\left( \frac{\theta}{\sqrt{1-s^2}} \right) \right) \right) \tag{20}$$

Noting that the cosine factor has a stationary phase at s=0 [2], we expand it around s=0,

$$\frac{1}{3} + \Re\left[\int_{s=0}^{s=\epsilon} ds(1-s^2)\exp\left(i\theta\left(1+\frac{s^2}{2}\right)\right)\right] \tag{21}$$

where the $s^2$ term can be neglected. The remaining integral can be further approximated to be:

$$\frac{1}{3} + \Re\left[\int_{s=0}^{s=\infty} ds\exp\left(i\theta\left(1+\frac{s^2}{2}\right)\right)\right] \tag{22}$$

which can now be evaluated exactly:

$$\frac{1}{3} + \frac{\sqrt{\pi}}{\sqrt{2}\,\theta}\cos\left(\theta+\frac{\pi}{4}\right); \tag{23}$$

As shown in the main text, the analytical approximation is in excellent agreement with the exact numerical integration. All other terms are evaluated similarly.

---



\end{document}